\documentclass[aps,prd,preprint,preprintnumbers,nofootinbib]{revtex4}
\usepackage{graphicx}

\usepackage[utf8]{inputenc} 

\usepackage{amsmath}
\usepackage{amssymb}

\usepackage[usenames,dvipsnames]{color}

\definecolor{mightnightblue}{RGB}{25,25,112}
\usepackage{hyperref}
 \hypersetup{
     colorlinks=true,
     linkcolor= Black,
     citecolor=mightnightblue,
     urlcolor=mightnightblue
     }

\def\lfv{lepton flavour violation }
 \def\cpv{CP violation }

\def\21{$\mathrm{SU(2)_L \otimes U(1)_Y}$}
\def\lfv{lepton flavour violation }

\newcommand{\sm}{standard model }

\newcommand{\AddrAHEP}{AHEP Group, Institut de F\'{i}sica Corpuscular --
  C.S.I.C./Universitat de Val\`{e}ncia, Parc Cientific de Paterna.\\
  C/Catedratico Jos\'e Beltr\'an, 2 E-46980 Paterna (Val\`{e}ncia) - SPAIN}

\newcommand{\Cinvestav}{Departamento de F\'{\i}sica, Centro de
  Investigaci{\'o}n y de Estudios Avanzados del IPN\\ Apdo. Postal
  14-740 07000 Mexico, DF, Mexico}
  
\newcommand{\Virginia}{Center for Neutrino Physics, Virginia Tech,
  Blacksburg, VA 24061, USA}
  
\begin{document}


\title{Probing CP violation with non-unitary mixing \\ in
 long-baseline neutrino oscillation experiments: \\DUNE as a case study}

\author{F. J. Escrihuela~$^1$}\email{franesfe@alumni.uv.es}
\author{D. V. Forero~$^2$}\email{dvanegas@vt.edu}
\author{O. G. Miranda~$^3$}\email{omr@fis.cinvestav.mx}
\author{M. T\'ortola~$^1$}\email{mariam@ific.uv.es}
\author{J. W. F. Valle~$^1$} \email{valle@ific.uv.es, URL:
  http://astroparticles.es/} 
\affiliation{$^1$~\AddrAHEP}
\affiliation{$^2$~\Virginia}
\affiliation{$^3$~\Cinvestav}

\begin{abstract}
  When neutrino masses arise from the exchange of neutral heavy
  leptons, as in most seesaw schemes, the effective lepton mixing
  matrix $N$ describing neutrino propagation is non-unitary, hence
  neutrinos are not exactly orthonormal.
  New CP violation phases appear in $N$ that could be confused with
  the standard phase $\delta_{\text{CP}}$ characterizing the three neutrino
  paradigm.
  We study the potential of the long-baseline neutrino experiment DUNE
  in probing CP violation induced by the standard CP phase in the
  presence of non-unitarity.
  In order to accomplish this we develop our previous formalism, so as
  to take into account the neutrino interactions with the medium,
  important in long baseline experiments such as DUNE.
  We find that the expected CP sensitivity of DUNE is somewhat
  degraded with respect to that characterizing the standard unitary
  case. However the effect is weaker than might have been expected
  thanks mainly to the wide neutrino beam.
  We also investigate the sensitivity of DUNE to the parameters
  characterizing non-unitarity. In this case we find that there is no
  improvement expected with respect to the current situation, unless
  the near detector setup is revamped.
 
  \end{abstract}


\maketitle
\section{Introduction}

Following the celebrated discovery of neutrino
oscillations~\cite{Kajita:2016cak,McDonald:2016ixn} subsequent
accelerator and reactor studies have brought neutrino physics to the
mature phase of precision studies.
Sensitive laboratory oscillation studies not only play a key role in
confirming the neutrino oscillation hypothesis, but also rule out
exotic solutions, establishing the robustness of the simplest three
neutrino paradigm.

Given its importance, more than ever it has become relevant to
critically assess with improved sensitivity the robustness of the
determination of the three-neutrino oscillation parameters within
recent and current
studies~\cite{Adamson:2016tbq,An:2016ses,Kim:2016yvm,An:2012eh,Ahn:2012nd,Abe:2012gx,Abe:2012tg,Abe:2015awa}
as well as future experiments~\cite{Acciarri:2015uup}.
This includes the scrutiny of the uncertainties associated with
neutrino fluxes, propagation and interactions. These may arise, for
example, from
helioseismology~\cite{christensen-dalsgaard:2002ur,Balantekin:2013tqa},
solar chemical composition and solar fusion
reactions~\cite{Adelberger:2010qa}, density fluctuations deep within
the Sun~\cite{balantekin:1996pp,nunokawa:1996qu} as well as magnetic
fields in the
radiative~\cite{burgess:2003fj,burgess:2003su,burgess:2002we} and
convective
zones~\cite{miranda:2000bi,barranco:2002te,Miranda:2003yh,Miranda:2004nz}.
On the other hand the subleading role of neutrino non-standard
  interactions upon oscillations has been considered in various
  contexts and can also bring new sources of CP
  violation~\cite{fornengo:2000sr,GonzalezGarcia:2001mp,Miranda:2004nb,deGouvea:2015ndi,Forero:2016cmb}.
These issues have been widely explored, so here we focus on the impact
of non-unitarity of the lepton mixing matrix upon neutrino propagation
and the resulting expected sensitivities on the three--neutrino CP
phase
determination~\cite{Goswami:2008mi,Escrihuela:2015wra,Miranda:2016wdr}.

Non-unitarity of the lepton mixing matrix constitutes a most generic
feature of schemes where neutrino masses arise from the exchange of
fermionic messengers \cite{valle:1987gv,Miranda:2016ptb} such as the
type-I seesaw
mechanism~\cite{GellMann:1980vs,Yanagida:1979as,mohapatra:1980ia,Schechter:1980gr}.
Indeed there is a large class of low-scale variants of the seesaw
mechanism, such as inverse and linear
seesaw~\cite{Mohapatra:1986bd,gonzalezgarcia:1989rw,Akhmedov:1995vm,Malinsky:2005bi},
where these right-handed neutrino messengers are not-so-heavy, as
their masses could lie within reach of the LHC experiments.
In this case one expects sizeable departures from unitarity in the
lepton mixing matrix characterizing the light neutrino
sector~\cite{Goswami:2008mi,Escrihuela:2015wra}. This brings in CP
violation associated to the messenger sector into the physics
describing the propagation of the light
neutrinos~\cite{Miranda:2016ptb}.
The presence of unitarity violation makes it difficult to extract
reliable information on leptonic \cpv and, indeed, first quantitative
studies indicate the existence of a potentially serious ambiguity in
probing CP violation in neutrino oscillations in such
case~\cite{Miranda:2016wdr}.
As a result, dedicated leptonic CP violation studies taking into
account the non-unitarity of the lepton mixing matrix will be
necessary. Such studies can shed light on the seesaw scale, and
thereby provide valuable insight on the scale of new physics
responsible for neutrino mass generation.

In this paper we focus on the possible ambiguities in the CP phase
determination for the upcoming DUNE experiment, including matter
effects in a consistent way. The paper is organized as follows.
In order to set up the framework in
section~\ref{sec:prel-updat-bounds} we compile and update the bounds
on the relevant parameters. These follow, for instance, from weak
universality tests and short--distance neutrino oscillation searches. 
In section~\ref{sec:non-unit-oscill} we discuss
the neutrino effective matter potential in the presence of
non-unitarity and present the corresponding results for the
oscillation probabilities in matter.
In section~\ref{sec:nuDUNE} we study the sensitivity of the DUNE
experiment to non-unitary neutrino mixing.  First we discuss the
determination of the standard three-neutrino CP phase
$\delta_{\text{CP}}$ and the possible confusion with the seesaw phase.
Finally we analyze the potential capability of DUNE in further
constraining the non-unitarity of the light neutrino mixing matrix.
We find that, although the potential to probe CP violation is
  somewhat degraded with respect to the unitary case, the effect is weaker than
  expected~\cite{Miranda:2016wdr} thanks to the good statistics
  expected and the relatively wide band neutrino beam at DUNE.
Further discussion and conclusions are summarized in
section~\ref{sec:conclusions}.

\section{preliminaries: prior constraints }
\label{sec:prel-updat-bounds}

Within a large variety of seesaw schemes the lepton mixing matrix
describing the propagation of the light neutrinos is effectively
non-unitary, hence these neutrino states are not exactly
orthonormal~\cite{valle:1987gv}.
The description of this situation can be readily obtained by
truncating the symmetrical parametrization of the full rectangular
lepton mixing matrix characterizing general seesaw schemes, first
given in~\cite{Schechter:1980gr}. The resulting form can be written
as~\cite{Escrihuela:2015wra}
 \begin{equation}
   N
 =
   N^{NP} U
 =
 \left\lgroup
 \begin{array}{ccc} 
 \alpha_{11} & 0 & 0\\
 \alpha_{21} & \alpha_{22} & 0\\
 \alpha_{31} & \alpha_{32} & \alpha_{33}
 \end{array}
 \right\rgroup U \,,
 \label{eq:NU}
 \end{equation}
 where $U$ is the conventional unitary mixing matrix describing
 neutrino propagation in the standard case, and the pre-factor
 parametrizes the deviations from unitarity.  This convenient
 description is general and holds for any number of extra neutrino
 states~\cite{Schechter:1980gr,Rodejohann:2011vc}.
 It involves three real parameters $\alpha_{11},\alpha_{22}$ and
 $\alpha_{33}$ (all close to one) plus three small complex parameters
 $\alpha_{21},\alpha_{31},\alpha_{32}$ containing extra CP violation.
The resulting form provides the most general framework to describe
neutrino oscillations relaxing the unitarity approximation.

In order to set the stage for our analysis we first give a brief
review on the constraints on non-unitarity parameters.  In what follows
we update the discussion given
in~\cite{Escrihuela:2015wra,Miranda:2016ptb}, e.g. by including recent
results for observables coming from pion decay
studies~\cite{Aguilar-Arevalo:2015cdf}. We also discuss the interplay,
as well as the complementarity, of various ``prior'' restrictions with
what can be learned by direct neutrino studies.
The bottom-line of our discussion will be that few of these
constraints are of general validity, most are model-dependent.

\subsection{Weak interaction without universality: formalism}
\label{sec:weak-inter-with}

Here we show how, from general considerations, the constraints from
weak no-universality translate into restrictions on non-diagonal
$\alpha_{ij}$ parameters.
In order to see this, we consider the parametrization of the non-unitary
lepton mixing matrix in Eq.~\ref{eq:NU}.  The diagonal entries of the
pre-factor matrix are given as a product of
cosines~\cite{Escrihuela:2015wra}:
\begin{equation}
\alpha_{11} \: = \: c_{1\, n}\: c_{1\,n-1}c_{1\, n-2}\ldots c_{14} ,
\end{equation}
 while the non-diagonal parameters are expressed as~\cite{Escrihuela:2015wra}:
\begin{eqnarray}
\alpha_{21} \: &=& \:
    c_{2\, n}\: c_{\,2n-1}\ldots c_{2\, 5}\:{\eta}_{24}\bar{\eta}_{14}\: +\: 
    c_{2\, n}\: \ldots c_{2\, 6}\:{\eta}_{25}\bar{\eta}_{15}\:c_{14} \nonumber\\ 
      &+&\:  \ldots\:+ {\eta}_{2n}\bar{\eta}_{1n}\:c_{1n-1}\:c_{1n-2}\:\ldots\:c_{14} 
\end{eqnarray}
with the phase factors ${\eta}_{ij}=e^{-i\phi_{ij}}\,\sin\theta_{ij}$
and
$\bar{\eta}_{ij}=-e^{i\phi_{ij}}\,\sin\theta_{ij}$~\cite{Schechter:1980gr}. 
Since the ``heavy'' iso-singlet admixture is assumed to be small,
within the framework of seesaw schemes, as well as from
experimental evidence~\cite{Astier:2003gs}, we now treat unitarity
violation as a perturbation, making use of an small-angle expansion in
$\theta_{i\beta}$, with $\beta > 3$, so that
\begin{equation}
\alpha^2_{11} \: \simeq 
\: 1-\sum^N_{i=4}\theta^2_{1\,i}\:.
\end{equation}
 On the other hand one can show that
\begin{eqnarray}
\alpha_{21} \: &\simeq& \:
-\theta_{24}\theta_{14}e^{-i(\phi_{24}-\phi_{14})}
    \: -\: 
    \theta_{25}\theta_{15}e^{-i(\phi_{25}-\phi_{15})} 
    \ldots - 
    \theta_{2n}\theta_{1n}e^{-i(\phi_{2n}-\phi_{1n})} 
\end{eqnarray}
so that
\begin{eqnarray}
|\alpha_{21}|^2 
\: &\simeq& \:
|  \theta_{24}\theta_{14}e^{-i(\phi_{24}-\phi_{14})}
    \: +\: 
    \theta_{25}\theta_{15}e^{-i(\phi_{25}-\phi_{15})} 
    \ldots: +
    \theta_{2n}\theta_{1n}e^{-i(\phi_{2n}-\phi_{1n})} |^2  \nonumber \\  
    &\leq& \sum^N_{i=4} |\theta_{2i}\theta_{1i}e^{-i(\phi_{2i}-\phi_{1i})} |^2 
     ~=~ \sum^N_{i=4} \theta^2_{2i}\theta^2_{1i} \nonumber
\end{eqnarray}
Now from the triangle inequality relation one can write
\begin{equation}
    \sum^N_{i=4} \theta^2_{2i}\theta^2_{1i} \leq 
    \left(\sum^N_{i=4}\theta^2_{2\,i} \right) 
    \left(\sum^N_{i=4}\theta^2_{1\,i} \right)
\end{equation}
which implies the relation
\begin{equation}
|\alpha_{21}| \le \sqrt{(1-\alpha_{11}^2)(1-\alpha_{22}^2)} \label{eq:CScond21} \\
\end{equation}
and similar relations will hold for $\alpha_{31}$ and $\alpha_{32}$, namely,
\begin{eqnarray}
|\alpha_{31}| &\le& \sqrt{(1-\alpha_{11}^2)(1-\alpha_{33}^2)}  \label{eq:CScond}\\ 
|\alpha_{32}| &\le& \sqrt{(1-\alpha_{22}^2)(1-\alpha_{33}^2)} \nonumber
\end{eqnarray}
One sees that in the limit of small heavy singlet messenger admixture
one has,
$$\alpha_{ii} \sim 1~~~\mathrm{and}~~~ \alpha_{ij} \ll1.$$
These relations imply additional restrictions on non-diagonal entries
coming from constraints on the diagonal ones.
In the next section we will see that bounds on diagonal entries are
relatively strong reinforcing the bounds on non-diagonal ones.
This also implies that \lfv and \cpv rates in the charged sector are
constrained mainly by universality restrictions, not by the smallness
of neutrino masses themselves.
This important observation has previously been made in a number of
papers and
reviews~\cite{bernabeu:1987gr,gonzalez-garcia:1992be,rius:1989gk}.
In the next subsection we compile bounds from universality as well as
from the relevant neutrino oscillation experiments.

\subsection{Universality constraints}
\label{sec:univ-constr}

The non-unitarity of the light neutrino mixing matrix can be
constrained by several observables related to weak universality.
\begin{itemize}
\item {\underline{\bf CKM unitarity}}
  
  As has been widely discussed in the literature~\cite{ Gronau:1984ct,
    Langacker:1988ur, GonzalezGarcia:1990fb, Nardi:1994iv,
    Abada:2012mc, Atre:2009rg, Abada:2013aba, Abada:2014nwa,
    Antusch:2014woa}, the comparison of measurements of muon and beta
  decay rates can constrain the non-unitarity of the neutrino mixing
  matrix.
  For example, the Fermi constant value for muon and beta decay will
  be proportional to different non-unitary parameter combinations:
\begin{equation}\label{eq1}
G_{\mu}=G_F\, \sqrt{(NN^\dagger)_{11}(NN^\dagger)_{22}}\, 
= G_F\, \sqrt{\alpha_{11}^2(\alpha_{22}^2+|\alpha_{21}|^2)},
\end{equation}
and
\begin{equation}\label{eq1A}
G_{\beta}=G_F\, \sqrt{(NN^\dagger)_{11}}\, 
= G_F\, \sqrt{\alpha_{11}^2}.
\end{equation}
This will imply that the CKM elements $V_{ud}$ and $V_{us}$,
proportional to the Fermi constant $G_{\mu}$, should be corrected by
the corresponding factor and expressed
as~\cite{Langacker:1988ur,GonzalezGarcia:1990fb,Nardi:1994iv}:
\begin{equation}\label{eq3}
\sum_{i=1}^{3} |V_{ui}|^2=
\left(\frac{G_\beta}{G_\mu}\right)^2 =
\left(\frac{G_F\sqrt{(NN^\dagger)_{11}}}{G_F\sqrt{(NN^\dagger)_{11}(NN^\dagger)_{22}}}\right)^2
=\frac{1}{(NN^\dagger)_{22}},
\end{equation}
The experimental value of this expression is given by~\cite{Olive:2016xmw}:
\begin{equation}\label{eq11}
\sum_{i=1}^{3} |V_{ui}|^2=\frac{1}{\alpha_{22}^2 + |\alpha_{21}|^2}
=0.9999\pm 0.0006,
\end{equation}

\item {\underline{\bf W mass measurements}} 

The mass of the $W$ boson, $M_W$, is related with the values of the weak mixing angle, $s_W$, and the
  Fermi constant. Including radiative corrections, in the On-Shell
  renormalization scheme, this relation can be written
  as~\cite{Olive:2016xmw}:
\begin{equation}
  M_{W} = \frac{A_{0}}{s_{W} (1-\Delta r)^{1/2}}
\, ,
\end{equation}
with
\begin{eqnarray}
A_{0} &=& \left(\frac{\pi \alpha}{\sqrt{2}G_{F}}\right)^{1/2} 
\, , \\
s^2_{W} &=&  0.22336 \pm
          0.00010 \, , \\
  \Delta r &=& 0.03648 \pm 0.00031 \, ,
\end{eqnarray}
where $\Delta r$ includes the radiative corrections relating $\alpha$,
$\alpha(M_{Z})$, $G_F$, $M_W$ and $M_Z$. In the non-unitary case, the
Fermi constant should take into account the corresponding corrections
and the prescription for $A_0$ will be:

\begin{equation}
A_{0} =
\left(\frac{\pi \, \alpha \, \sqrt{\alpha^{2}_{11}(\alpha^{2}_{22} + |\alpha^{2}_{21}|)}}{\sqrt{2}G_{\mu}}\right)^{1/2}
\, .
\end{equation}

\item {\underline{\bf semileptonic weak decays}} 

  The couplings between leptons and gauge bosons are dictated by gauge
  symmetry.  For the standard case of lepton unitarity these are
  flavor independent. This feature is no longer true in the presence
  of non-unitarity. As a result, the ratio between two different
  semileptonic decay rates would constrain non-unitarity
  parameters. For example, for the case of pion decay we
  have~\cite{Abada:2012mc}:
\begin{equation}
R_{\pi} = \frac{\Gamma (\pi^+ \to e^+\nu)}{\Gamma (\pi^+ \to \mu^+\nu)}= 
\frac{(NN^\dagger)_{11}}{(NN^\dagger)_{22}} =
\frac{\alpha_{11}^2}{\alpha_{22}^2 + |\alpha_{21}|^2}.
\end{equation}
Here we include the updated measurement from
Ref.~\cite{Aguilar-Arevalo:2015cdf} and theoretical prediction in
Refs.~\cite{Cirigliano:2007xi,Bryman:2011zz}:
\begin{equation}\label{eq10}
r_\pi = \frac{R_\pi}{R_\pi^{SM}} =
\frac{(1.2344\pm 0.0029)\times 10^{-4}}{(1.2352\pm 0.0002)\times 10^{-4}}=
0.9994\pm0.0030 .
\end{equation} 
Notice that this constraint is more restrictive than the previously
reported value, $r_\pi=0.9956\pm0.0040$~\cite{Czapek:1993kc}.
One also has the corresponding bound from Kaon-decay \cite{Abada:2012mc}
\begin{equation}\label{eq10}
r_K= \frac{R_{K}}{R^{SM}_{K}} =
\frac{(NN^\dagger)_{11}}{(NN^\dagger)_{22}} =
\frac{\alpha_{11}^2}{\alpha_{22}^2 + |\alpha_{21}|^2} =
\frac{(2.488\pm 0.010)\times 10^{-5}}{(2.477\pm 0.001)\times 10^{-5}}=1.004\pm0.010.
\end{equation}
However, this limit does not play a significant role, since the pion decay
measurements are more restrictive.

\item {\underline{\bf $\mu-\tau$ universality }} 

  Likewise, for the case of $\mu-\tau$ universality there are
  restrictions that follow from the ratio of the decay of the meson
   ($\pi^{-}$ or $K^{-}$) to
  a muon plus a muon neutrino, or from the tau decay to a meson
 and a tau neutrino~\cite{Nardi:1994iv,
    Escrihuela:2015wra, Pich:2013lsa}:

\begin{equation}
R_{\tau / P} = \frac{\Gamma (\tau^- \to P^-\nu_{\tau})}
   {\Gamma (P^- \to \mu^+\nu_{\mu})} \propto 
{\left|\frac{g_{\tau}}{g_{\mu}}\right|}^{2} = 
\frac{\alpha_{33}^2 + |\alpha_{32}|^2 + |\alpha_{31}|^2
}{\alpha_{22}^2 + |\alpha_{21}|^2}, 
\end{equation}
where $P^{-}$ stands for either $\pi^{-}$ or $K^{-}$ mesons. Several
ratios can be considered and included in the analysis.  In particular
we have considered the results reported in Ref.~\cite{Pich:2013lsa}.\\

\item {\underline{\bf $e-\tau$ universality }} 

  On the other hand, for the $e$ - $\tau$ sector, we have considered
  only pure leptonic decays as well as direct leptonic decays of $W$
  boson, which lead to
\begin{equation}
{\left|\frac{g_e}{g_{\tau}}\right|}^2 = \frac{\alpha_{11}^2}{\alpha_{33}^2 + |\alpha_{32}|^2 + |\alpha_{31}|^2
} \, .
\label{eq:g_e_tau}
\end{equation}
The value of $|g_{e}/g_{\tau}|$ ratio for each process was presented
in Ref.~\cite{Pich:2013lsa}.

\item \underline{Invisible Z decay width}

  Non-unitarity can affect the neutral current couplings. As noted
  in~\cite{Schechter:1980gr}, these are no longer ``trivial'' as in
  the \sm since the couplings of light neutrinos to the Z-boson can be
  non-diagonal in the mass basis.  Moreover the diagonal coupling
  strengths are smaller than in the \sm thereby decreasing the
  invisible Z width, well measured at LEP and reported to be slightly
  smaller than three ($2.9840\pm0.0082$)~\cite{ALEPH:2005ab}. However,
  neutral currents have a more complex structure that will depend both
  on the values of the $\alpha$ parameters as well as on the values of
  the three by three matrix $U^{3\times3}$. Given this complexity and the
  quadratic dependence on  the $\alpha$, it is
  safe not to include this observable into the analysis.
\end{itemize}

Concerning searches for lepton flavor and CP violating processes we
notice that these do not give us any independent robust constraint on
unitarity violation. Indeed, such processes may proceed in the absence
of neutrino mass and are only restricted by weak universality
tests~\cite{bernabeu:1987gr,branco:1989bn,rius:1989gk}.

\begin{table}
\begin{tabular}{ccc|cc}
\hline \hline
\multicolumn{3}{c}{One parameter} & \multicolumn{2}{|c}{All parameters}\\
\multicolumn{3}{c}{(1 d.o.f.)} & \multicolumn{2}{|c}{(6 d.o.f.)} \\
\hline 
 & $90$\% C.L. &  $3\sigma$ & $90$\% C.L. & $3\sigma$ \\
 \hline \hline
\multicolumn{5}{c}{Neutrinos + charged leptons} \\
 \hline \hline
$\alpha_{11}>$    &$0.9974$ &$0.9963$&$0.9961$ &$0.9952$         \\ 
$\alpha_{22}>$    &$0.9994$&$0.9991$&$0.9990$&$0.9987$          \\ 
$\alpha_{33} >$    & $0.9988$ & $0.9976$&  $0.9973$ & $0.9961$ \\ 
$|\alpha_{21}|<$  &  $\;1.7\times10^{-3}\;$ & $\;2.5\times10^{-3}\;$&  $\;2.6\times10^{-3}\;$ & $\;4.0\times10^{-3}\;$  \\ 
$|\alpha_{31}|<$  &  $2.0\times10^{-3}$ & $4.4\times10^{-3}$&  $5.0\times10^{-3}$ & $7.0\times10^{-3}$ \\ 
$|\alpha_{32}|<$  &  $1.1\times10^{-3}$ & $2.0\times10^{-3}$&  $2.4\times10^{-3}$ & $3.4\times10^{-3}$  \\ 
\hline \hline
\multicolumn{5}{c}{Neutrinos only} \\
\hline \hline
$\alpha_{11}>$   & $0.98$& $0.95$   & $0.96$ & $0.93$ \\
$\alpha_{22}>$   & $0.99$ & $0.96$ & $0.97$ &  $0.95$ \\
$\alpha_{33}>$   & $0.93$ & $0.76$ & $0.79$ &  $0.61$ \\
$|\alpha_{21}|<$  &  $1.0 \times10^{-2}$  &  $2.6\times10^{-2}$  &
                                $2.4\times10^{-2}$  & $3.6 \times10^{-2}$    \\ 
$|\alpha_{31}|<$  & $4.2 \times10^{-2}$  & $9.8\times10^{-2}$  &
                                $9.0\times10^{-2}$  &   $1.3\times10^{-1}$   \\ 
$|\alpha_{32}|<$  &  $9.8 \times10^{-3}$  & $1.7\times10^{-2}$ &
                                 $1.6\times10^{-2}$  &  $2. 1 \times10^{-2}$  \\ 
\hline \hline
\end{tabular}
\caption{\label{tab:bounds} Bounds on the magnitudes of the non-unitarity parameters at $90$\%\,C.L. and
    $3\sigma$ (for 1 and 6 d.o.f.). Upper table: constraints coming from neutrinos and charged leptons.  Lower part:
    constraints derived from direct neutrino oscillation searches~\cite{Parke:2015goa, Astier:2001yj,Astier:2003gs}.}
\end{table}
\vspace{1cm}

\subsection{Neutrino oscillation constraints}

Direct constraints on the non-diagonal elements of the $N$ matrix come
from the so-called zero distance effect in the conversion
probability~\cite{valle:1987gv}.  For example, the conversion
probability from muon to electron neutrinos can be written
as~\cite{Escrihuela:2015wra},
\begin{equation}
P_{\mu e} \simeq 
 (\alpha_{11}\alpha_{22})^2 P^{3\times3}_{\mu e}
+  \alpha_{11}^2 \alpha_{22}|\alpha_{21}| P^{I}_{\mu e} 
+ \alpha_{11}^2|\alpha_{21}|^2 ,
\label{eq:Pmue}
\end{equation}
after neglecting cubic products of the small parameters $\alpha_{21}$,
$\sin\theta_{13}$ and $\Delta m^2_{21}$. Here, $P^{3\times3}_{\mu e}$
stands for the standard conversion probability in the unitary case,
while the interference probability term $P^{I}_{\mu e}$ depends on the
non-unitarity parameters, including an additional $CP$ phase. Finally,
the last term in this expression is a constant factor, independent of
the distance travelled by the neutrino and its energy. Therefore, any
neutrino appearance experiment in the $\nu_\mu \to \nu_e$ channel
would be sensitive to this zero distance (0d) contribution:
\begin{equation}
P^{0d}_{\mu e} = \alpha_{11}^2 |\alpha_{21}|^2 .
\end{equation}
 There is a similar expression for the conversion probability in the  $\nu_e \to \nu_\tau$ and 
$\nu_\mu \to \nu_\tau$  channels.  In
the latter case, the oscillation probability formula is slightly more
complicated, but at leading order in the non-unitary parameters, one
can approximate both zero-distance appearance probabilities by 
 \begin{eqnarray}
P^{0d}_{e \tau} & = \alpha_{11}^2 |\alpha_{31}|^2  \, , \\
P^{0d}_{\mu \tau} & \simeq \alpha_{22}^2 |\alpha_{32}|^2 \, .
\end{eqnarray}
We have used these expressions to obtain direct constraints on the 
parameters $|\alpha_{21}|$, $|\alpha_{31}|$ and $|\alpha_{32}|$ using
the negative searches from the  NOMAD and CHORUS short-baseline experiments. 
NOMAD~\cite{Astier:2003gs,Astier:2001yj}  reported
limits on the search for $\nu_{\mu} \to \nu_{e}$  as well as $\nu_{\mu} \to \nu_{\tau}$ 
oscillations in a predominantly $\nu_{\mu}$ neutrino beam produced by the SPS at
CERN, while CHORUS~\cite{Eskut:1997ar,Eskut:2007rn} used the same beam 
to search for $\nu_{\mu} \to \nu_{\tau}$ oscillations. 
Additionally, from the contamination of electron neutrinos in the beam, they were also able to constrain 
the oscillation channel $\nu_e \to \nu_{\tau}$. 
The stronger bounds at 90\% C.L. from these experiments, obtained by NOMAD, can be summarized
as:
\begin{eqnarray}
P^{0d}_{\mu e} & < &7.0\times 10^{-4}  \nonumber \\
P^{0d}_{\mu \tau} &  <  &1.6\times 10^{-4}  \\
P^{0d}_{e \tau} & < & 0.74\times 10^{-2} \nonumber
%
 \label{eq:0d-bounds}
\end{eqnarray}
Similar constraints can also be obtained from the NuTeV
  data~\cite{Avvakumov:2002jj}. 
Note that, in addition to the short-baseline experiments 
 discussed above, there are also nontrivial constraints
  arising from medium and long-baseline experiments in combination with atmospheric and 
  solar neutrino data~\cite{Parke:2015goa,Qian:2013ora}.
  For maximal values of the diagonal parameters $\alpha_{ii}$, one can
  summarize the bounds obtained in~\cite{Parke:2015goa} in terms of
  3$\sigma$ limits on the non-diagonal parameters:
\begin{eqnarray}
|\alpha_{21}| <  &                     0.03    \nonumber \\
|\alpha_{31}| <  &                     0.11      \\
|\alpha_{32}| <   &                      0.12     \nonumber
\label{eq:osc-bounds}
\end{eqnarray}

 We stress that the above constraints coming from neutrino
  oscillations are independent of the mass scale of the heavy
neutrinos. Therefore, they can be used to constrain the non-unitarity
of the lepton-mixing matrix independent on the heavy mass scale. These
are the only fully model-independent constraints. Therefore, such
neutrino-data-only bounds play a special role and for this reason have
been separated as the lower part in~Table~\ref{tab:bounds}.
This Table summarizes all the available non-unitarity bounds discussed
in this section.
Clearly, as seen from the upper part of Table~\ref{tab:bounds}, one
can see that universality tests provide strong constraints on the
diagonal parameters, $\alpha_{ii}$, that are very close to unity,
independently of the number of degrees of freedom considered.
In addition, one can also combine with the relations in
Eqs.~(\ref{eq:CScond21}) and~(\ref{eq:CScond}) in order to obtain
stronger constraints on the non-diagonal $\alpha$ parameters.
  Indeed, by combining universality bounds with these relations one
  finds that the constraints on the non-diagonal parameters are of
  order $10^{-3}$.

  However we note that these limits are all derived from charged
  current induced processes under the restrictive assumption that
  there is no new physics other than that of non-unitary mixing.
As an example, we note that the presence of neutrino-scalar
    Yukawa interactions, absent in the \sm but present in models with
    extra Higgs bosons, such as multi-Higgs schemes
    (e.g. incorporating flavor symmetries), would potentially avoid
    these bounds. Likewise, the presence of right-handed charged
    current contributions expected within a left-right symmetric
    seesaw scheme would have the same effect.
    This happens if the extra scalar or vector-mediated contributions
    compensate the unitarity violation effect~\footnote{While direct
      search bounds for such charged mediators are rather stringent,
      one can still find ``fine-tuned'' funnels in parameter space
      which allow the situation envisaged here.}. Of course one may go
    beyond the above well-motivated assumptions and consider, for the
    sake of generality, the most general Lorentz structure for the
    charged weak interactions~\footnote{Such model-independent studies
      of the charged current weak interactions were given
      in~\cite{Mursula:1984zb,Fetscher:2008zz}.}.
    In such case these limits would be invalidated, leading us to
    regard them as fragile.
    In contrast, the constraints from neutrino experiments provide a direct restriction on the non-unitarity
    $\alpha$ parameters. These bounds are significantly less
    stringent, of the order of $10^{-2}$ for the non-diagonal $\alpha_{ij}$, and correspond to the lower
    entries in Table~\ref{tab:bounds}.

    Finally, there are also direct bounds from searches for neutral
    heavy leptons.
  These depend on the mass of the heavy neutrinos, and do not apply
  beyond the kinematical reach of the high energy experiments, such as
  LEP~\cite{Dittmar:1989yg,Akrawy:1990zq,abreu:1997pa} and
  LHC~\cite{Klinger}. All mass-dependent limits on light and heavy
  singlet neutrinos have been compiled in
  Refs.~\cite{Escrihuela:2015wra,Drewes:2015iva,Alekhin:2015byh,Miranda:2016ptb,Drewes:2016jae}.

\vskip2mm

In short, at this stage one may adopt two approaches:
\begin{itemize}
\item to use as reference the more restrictive bounds coming from
  charged current weak processes (upper part of Table
  \ref{tab:bounds})
\item to use as benchmarks bounds taken strictly from the neutrino
  sector (lower part of Table \ref{tab:bounds}).
\end{itemize}

While the top limits on the $\alpha$'s are stronger, they are not
robust enough for our purposes, so we would recommend to focus 
on the most direct constraints coming from the bottom part of Table
\ref{tab:bounds}).
In any case in our simulations for the DUNE experiment in order to
evaluate its potential in probing leptonic CP violation in the
presence of unitarity violation we include as benchmark values not
only the conservative, but also the model-dependent bounds, for
comparison.
The bottom-line is that the DUNE experiment will have the potential of
providing independent and robust probes of neutrino properties beyond
standard oscillations, properties which can not be probed otherwise in
a model-independent way.

\section{Non-unitary oscillation probabilities in matter}
\label{sec:non-unit-oscill}

In \cite{Escrihuela:2015wra,Miranda:2016wdr} we have given the
analytic expressions for the neutrino oscillation probabilities in
vacuum~\footnote{See Refs.~\cite{Antusch:2006vwa,
    Fernandez-Martinez:2016lgt}, where a different form for the
  non-unitary neutrino mixing matrix is used.}.  This approach is
valid to study oscillation experiments where matter effects are not
very relevant.
However, in order to obtain direct sensitivities on the non-unitarity
of the lepton mixing matrix from upcoming long-baseline neutrino
oscillation experiments such as DUNE or NO$\nu$A, one must have a
consistent way to describe matter effects appropriate to this
situation.
In order to quantify the impact of non-unitary mixing in such
experiments, where matter effects are expected to play an important
role, one should take into account how the effective matter potential
for neutrinos gets modified in the presence of non-unitary
three-neutrino mixing.  We discuss this issue in the next subsection.

\subsection{Neutrino effective matter potential 
in the presence of non-unitarity }
\label{sec:neutr-effect-matt}

The standard derivation of the effective potential that neutrinos feel
when traversing a material medium assumes unitary mixing between the
light neutrino species~\footnote{We will assume a non-polarized
  neutral medium in the calculation of the effective matter
  potential}.
In order to derive the neutrino potential in matter for a model with
neutral heavy leptons, we note that the complete expression for the
neutrino in a flavor state will be given by 
\begin{equation}
\nu_{\alpha} = \sum_{i}^n K_{\alpha i} \nu_{i} \, ,
\label{eq:nu_field_rel}
\end{equation}
with the $\alpha$ subscript indicating flavor and $i$ mass
eigenstates.
\begin{figure}
\begin{center}
  \includegraphics[width=0.79\textwidth,angle=0]{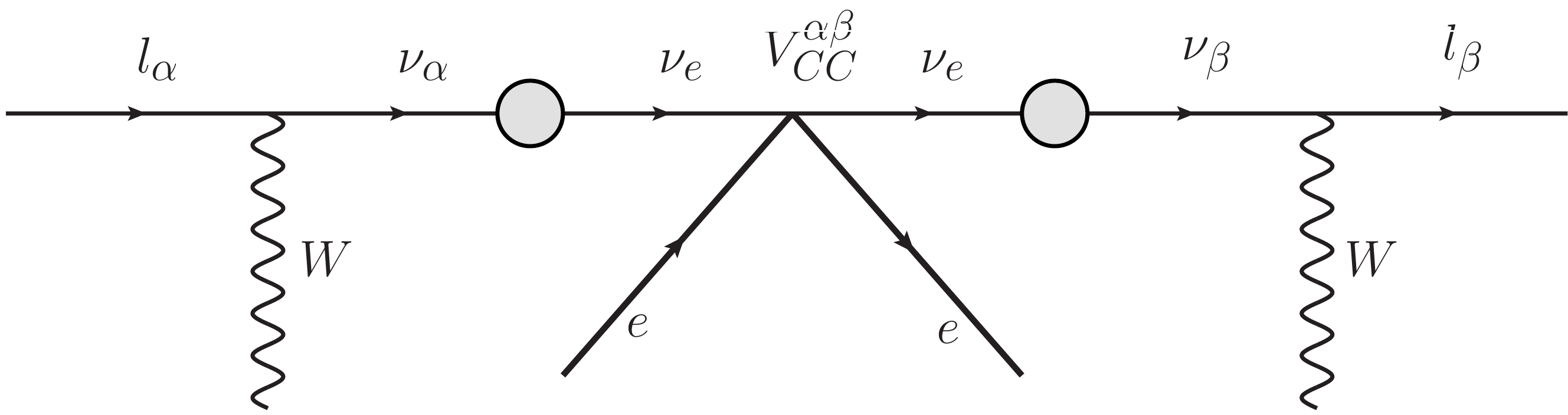}
  \caption{Feynman diagram illustrating the 
  matter potential  associated to the charged current~\cite{Wolfenstein:1977ue}.  }
\label{fig:Feyn_CC}
\end{center}
\end{figure}
Charged current matter effects in neutrino propagation are illustrated
in the Feynman-like diagram in Fig.~(\ref{fig:Feyn_CC}) and will be
proportional to
\begin{equation}
K_{\alpha i} K^{*}_{e i} K_{e j} K^{*}_{\beta j} =
(KK^{\dagger})_{\alpha e} \, (KK^{\dagger})_{e \beta}\, .
\label{eq:KK_rel}
\end{equation}
Therefore, the charged current potential will be given by 
\begin{equation}
V^{\alpha \beta}_{CC} =
 \sqrt{2} \,G_{F} N_{e} \left(KK^{\dagger}\right)_{\alpha e}
    \left(KK^{\dagger}\right)_{e \beta} \, .
\label{eq:V_cc_mass}
\end{equation}
where $N_e$ is the number density of electrons in the medium and $G_F$
is the Fermi constant.
However, the heavy states will not take part in a long baseline
neutrino oscillation set up. As a result the sum in
Eq.~(\ref{eq:nu_field_rel}) must be performed only up to the third
mass eigenstate.
Therefore, effectively, one has:
\begin{equation}
\nu_{\alpha} = \sum_{i}^3 K_{\alpha i} \nu_{i} =   \sum_{i}^3 N_{\alpha i} \nu_{i} \, ,
\label{eq:nu_rel_3}
\end{equation}
and the effective CC potential in the presence of non-unitarity will
be given by:  
\begin{equation}
V^{\alpha \beta}_{CC} =
 \sqrt{2} \,G_{F} N_{e} \left(NN^{\dagger}\right)_{\alpha e}
    \left(NN^{\dagger}\right)_{e \beta} \, .
\label{eq:V_cc_mass2}
\end{equation}
which is expressed in terms of the $\alpha$ parameters as:
\begin{equation}
(NN^\dagger)_{\alpha e}(NN^\dagger)_{e \beta}= \alpha_{11}^2
\left(
\begin{array}{ccc}
 \alpha _{11}^2 & \alpha _{11} \alpha _{21}^* & \alpha _{11} \alpha _{31}^* \\
 \alpha _{11} \alpha _{21} & |\alpha _{21}|^2 & \alpha _{21} \alpha _{31}^* \\
 \alpha _{11} \alpha _{31} & \alpha _{21}^* \alpha _{31} & |\alpha _{31}|^2 \\
\end{array}
\right)\, .
\label{eq:CC_alpha_corr}
\end{equation}
Clearly in the unitary limit ($\alpha_{ii} = 1$ and $\alpha_{ij}=0$),
one recovers the well-known Wolfenstein form for the effective CC
potential:
 \begin{equation}
V^{\alpha\beta}_{CC} = \sqrt{2}G_F N_e \delta_{\alpha e}\delta_{\beta e}
\label{eq:V_cc_SM}
\end{equation}

\begin{figure}[!t]
\begin{center}
  \includegraphics[width=0.79\textwidth,angle=0]{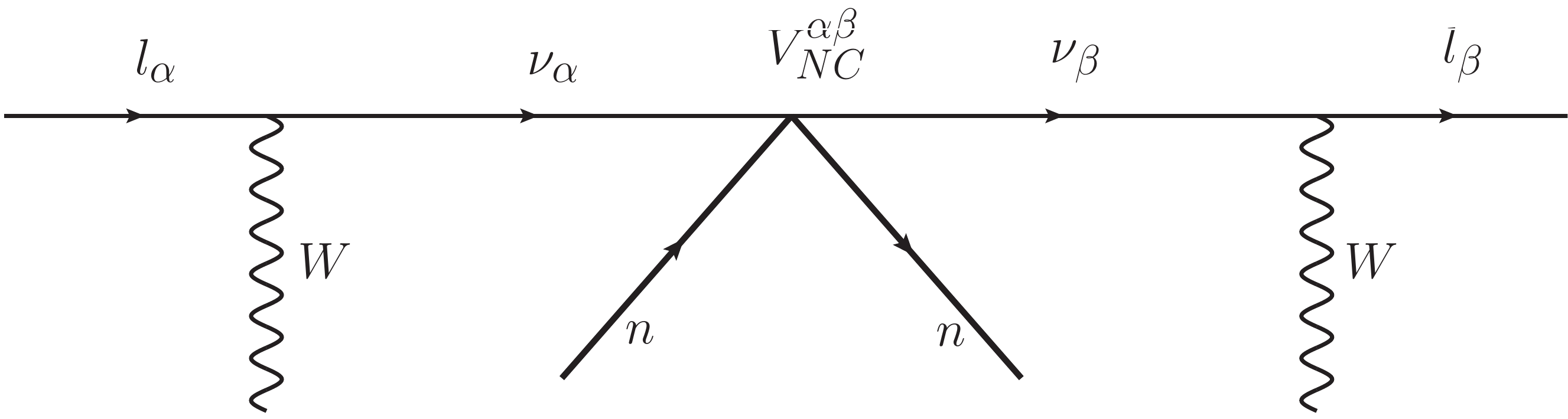}
  \caption{Feynman diagram illustrating the matter potential
    associated to the neutral current~\cite{valle:1987gv}.
  	     }
\label{fig:Feyn_NC}
\end{center}
\end{figure}

For the neutral current case we proceed in a similar way. Again we
consider the Feynman-like diagram of the NC process as described in
Fig. \ref{fig:Feyn_NC} and the neutral current potential is given by
\begin{equation}
V^{\alpha \beta}_{NC} =
   - \sum_{\rho} \frac{1}{\sqrt{2}}G_{F} N_{n} \left( KK^{\dagger}
    \right)_{\alpha \rho} \left( KK^{\dagger} \right)_{\rho \beta}\, .
\label{eq:V_NC}
\end{equation}
After truncating the rectangular $K$ matrix into the square matrix $N$,
we obtain that the NC contribution to the matter potential is given
by:
\begin{equation}
V^{\alpha \beta}_{NC}=-\sqrt{2}G_F \frac{N_n}{2}
\sum_{\rho}(NN^\dagger)_{\alpha \rho}(NN^\dagger)_{\rho
  \beta}=-\sqrt{2}G_F \frac{N_n}{2} \left[(NN^\dagger)^2
\right]_{\alpha \beta} \, ,
\label{eq:V_NC_corr}
\end{equation}
where the matrix product $(NN^\dagger)^2 $ at leading order in
  the non-diagonal $\alpha$'s is given by:
\begin{equation}\label{eq:matt-pot-app}
\left(
\begin{array}{ccc}
 \alpha_{11}^4 & \alpha_{11} \alpha^*_{21} \left(\alpha_{11}^2+\alpha_{22}^2\right) & \alpha_{11} \alpha^*_{31} \left(\alpha_{11}^2+\alpha_{33}^2\right) \\
 \alpha_{11} \alpha_{21} \left(\alpha_{11}^2+\alpha_{22}^2\right) & \alpha_{22}^4 & \alpha_{22} \alpha^*_{32} \left(\alpha_{22}^2+\alpha_{33}^2\right) \\
 \alpha_{11} \alpha_{31} \left(\alpha_{11}^2+\alpha_{33}^2\right) & \alpha_{22} \alpha_{32} \left(\alpha_{22}^2+\alpha _{33}^2\right) & \alpha_{33}^4 \\
\end{array}
\right)
\end{equation}
So, one sees how, starting with diagonal CC and NC potentials, due to
the non-unitarity one ends up in general with non-diagonal forms for
the effective matter potentials.

Notice that the non-unitarity parameters $\alpha_{31}$, $\alpha_{32}$
and $\alpha_{33}$, which do not enter in the expression of $P_{\mu e}$
in vacuum~\cite{Escrihuela:2015wra,Miranda:2016ptb}, do appear in the
calculation of $P_{\mu e}$ in matter due to the form of the effective
matter potential. 
The effect of the non-diagonal parameters $\alpha_{31}$ and
$\alpha_{32}$ is not as important as the role of $\alpha_{21}$.  The
$\alpha_{31}$ parameter enters linearly in the CC and NC potential in
the 13 entry.
Its effect will be analogous to that of the parameter
$\epsilon_{e\tau}$ in the case of non-standard interactions, so that
the resulting degeneracy with the reactor angle
$\theta_{13}$~\cite{Huber:2001de,Huber:2002bi} will imply a
deterioration of the sensitivity to CP
violation~\cite{Forero:2016cmb}.  In contrast, $\alpha_{32}$ will
enter only in the neutral current potential in the 23 entry and,
therefore, is expected to have a negligible impact.\\ [-0.2cm]

  Adding the two contributions to the effective potential in matter we will have~\footnote{Similar results have been obtained in Ref.\cite{Blennow:2016jkn}}:
\begin{equation}
V^{\alpha \beta} =  V^{\alpha \beta}_{CC} + V^{\alpha \beta}_{NC} = \sqrt{2} \,G_{F} N_{e} \left(NN^{\dagger}\right)_{\alpha e}
    \left(NN^{\dagger}\right)_{e \beta} 
 -\sqrt{2}G_F \frac{N_n}{2}
\sum_{\rho}(NN^\dagger)_{\alpha \rho}(NN^\dagger)_{\rho
  \beta}
    \label{eq:V_tot}
\end{equation}
where $\alpha$ and $\beta$ stands for the initial and final neutrino
flavor, respectively, and $\rho$ implies a sum over the three active
flavors. $N_e$ is the electron density in the medium while $N_n$ is
the neutron density.  
In  matrix form one has the following expression for the matter potential in the presence of non-unitarity:
\begin{equation}
V_{NU} =  \left(NN^{\dagger}\right)
\left[ \sqrt{2} \,G_{F} N_e 
\left(\begin{array}{ccc}  
1 & 0 & 0\\
0 & 0 &0 \\
0 & 0 & 0
\end{array}\right) - \frac{\sqrt{2}}{2} \,G_{F} N_n
 \left(\begin{array}{ccc}  
 1 & 0 & 0\\
0 & 1 &0 \\
0 & 0 & 1
\end{array}\right)\right] \left(NN^\dagger\right)    
\label{eq:V_tot2}
\end{equation}
leading to a very simple compact form
\begin{equation}
V_{NU} =  \left(NN^{\dagger}\right)
V_\mathrm{unitary}(NN^\dagger)~.  
\label{eq:V_tot3}
\end{equation}
Notice that, in contrast to the standard procedure used in the
three--neutrino unitary case, the contribution of the neutral current
potential can no longer be neglected when treating the non-unitary
case.

One can also see how to get this result from the truncation of the
$N\times N$ mixing matrix, $U$.  Therefore, the Hamiltonian in matter
in the flavour basis will be given by:
\begin{equation}
H_{NU} =  N
\left(\begin{array}{ccc}  
0 & 0 & 0\\
0 & \frac{\Delta m^2_{21}}{2E} &0 \\
0 & 0 &  \frac{\Delta m^2_{31}}{2E}
\end{array}\right) N^{\dagger} +
\left(NN^{\dagger}\right)
\left(\begin{array}{ccc}  
V_{cc} +V_{nc} & 0 & 0\\
0 & V_{nc} &0 \\
0 & 0 & V_{nc}
\end{array}\right) \left(NN^\dagger\right)    
\label{eq:V_tot2}
\end{equation}
with $V_{cc} = \sqrt{2} G_F N_e$ and
$V_{nc} = - \frac{\sqrt{2}}{2} G_F N_n$.

\section{ Non-unitary neutrino mixing in DUNE}
\label{sec:nuDUNE}

  Here we explore the expected sensitivities to the non-unitarity of
  the neutrino mixing matrix within the upcoming DUNE experiment.
  Previous studies have already considered the impact of non-unitarity
  upon the CP-phase sensitivity at T2K~\cite{Ge:2016xya}.
  Here we present a dedicated study for the DUNE experiment, whose
  longer baseline implies that matter effects are more relevant than
  for the cases of T2K and NOvA and therefore the formalism described
  above is crucial.
  
\subsection{DUNE simulation with non-unitary neutrino mixing}
\label{sec:dune-simul}

\begin{figure}
\begin{center}
\includegraphics[width=\textwidth,angle=0]{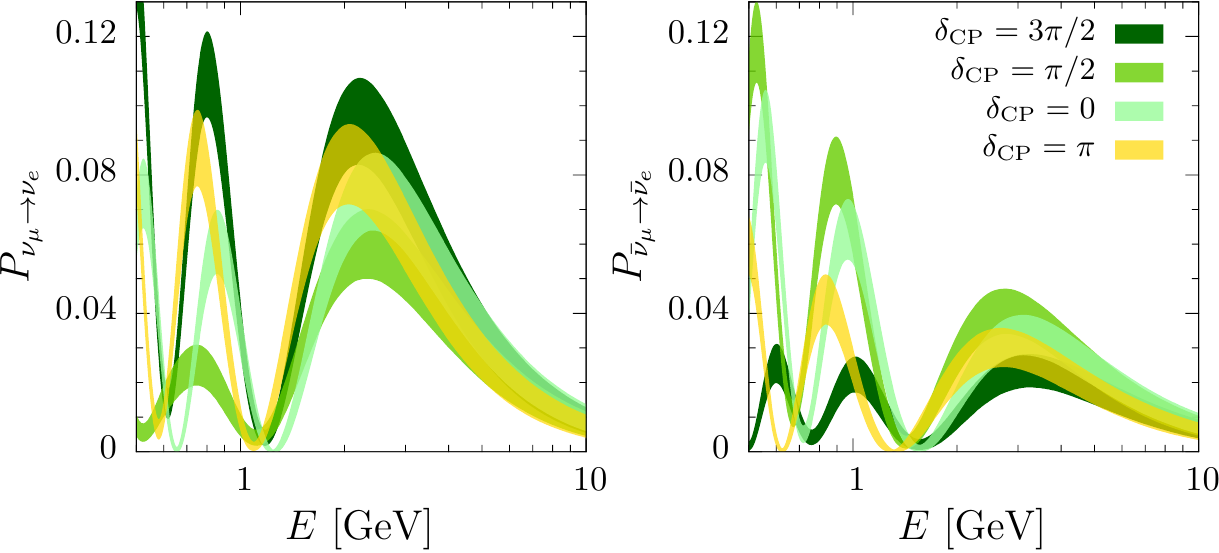}
\caption{Neutrino electron appearance probability in DUNE for the
  neutrino (left) and antineutrino (right panel) channel, with
  $|\alpha_{21}| = 0.02$ and $\phi_{21}$ free for fixed value of
  $\delta_{CP}$.}
\label{fig:dune}
\end{center}
\end{figure}

DUNE is a long-baseline neutrino oscillation experiment that will
measure neutrino oscillations over a broad energy range, from hundreds
of MeV to few tenths of GeV. This experiment will detect neutrinos and
anti-neutrinos produced in the NuMI beam line at Fermilab $1300$~km
away from the source, with relevant matter effects in the neutrino
propagation.
The effect of non-unitary neutrino mixing in the DUNE simulation will
modify the standard calculation of neutrino oscillation probabilities.
Besides the non-unitary neutrino mixing matrix, to calculate the
neutrino conversion probability in DUNE one must take into account the
modified matter potential affecting neutrino propagation through the
Earth, as discussed in the previous section.
Using the neutrino Hamiltonian with matter effects as given by
Eq.~(\ref{eq:V_tot2}), we have solved numerically the evolution
equation, obtaining the corresponding conversion probability from muon
to electron neutrino in the case of DUNE.
We illustrate the behavior of the modified neutrino appearance
probability $P_{\mu e}$ for the DUNE experiment in
Fig.~\ref{fig:dune}.
The left panel corresponds to neutrino probability and the right panel
to antineutrino probability.
Each band corresponds to a different value of the standard CP phase,
$\delta_{\text{CP}}$, while the width of the band is due to the variation
over the non-unitary phase $\phi_{21} = Arg(\alpha_{21})$.
The only deviation from unitarity in this calculation comes from the
$\alpha_{21}$ parameter, set to 0.02.
The overlap of the different bands indicates the presence of
degeneracies in the neutrino oscillation probability in DUNE.
This ambiguity, present at the probability level has already been
noticed in Ref.~\cite{Miranda:2016ptb,Miranda:2016wdr}
 
The DUNE experimental setup assumed for this analysis corresponds to a
$40$--ton liquid argon far detector with optimized neutrino fluxes,
cross sections, detector efficiency and energy resolution effects as
provided in the form of GLoBES~\cite{Huber:2004ka,Huber:2007ji} files
in Ref.~\cite{Alion:2016uaj}.
The calculation of the neutrino oscillation probabilities in the
presence of non-unitarity has been implemented in the GLoBES package
with an adequate modification of its probability engine.
In our analysis, we have used the spectral event information from the
four neutrino oscillation channels: electron (anti)neutrino appearance
and muon (anti)neutrino disappearance. To statistically quantify the
effect of the non-unitary lepton mixing parameters we have used the
usual $\chi^2$ definition adding penalties on the `unitary'
oscillation parameters $\theta_{ij}$ and $\Delta
m^2_{k1}$~\cite{Forero:2014bxa}. The relative error on these
parameters and the systematic uncertainties in the normalization of
signal and background for each oscillation channel, ranging from
$0.2\%$ to $20\%$ depending on the channel, were set to the values
given in Ref.~\cite{Alion:2016uaj}.

For the rest of this section we denote the three mixing angles
collectively as a vector
$\vec{\lambda}=\{\theta_{ij},\Delta m^2_{k1}\}$. We have included
penalties to the $\chi^2$ accounting for the allowed values of the
$\vec{\lambda}$ parameters. Likewise, we denote the non-unitarity
parameters in a compact form as
$\vec{\alpha}=\{\alpha_{ii},\alpha_{ij}\}$, including both their
diagonal and non-diagonal components. Note that we treat the CP phase
$\delta_{CP}$ separately.

\subsection{DUNE sensitivity to CP violation}
\label{sec:DUNE-CP}

\begin{figure}[!t]
\includegraphics[width=0.8\textwidth,angle=0]{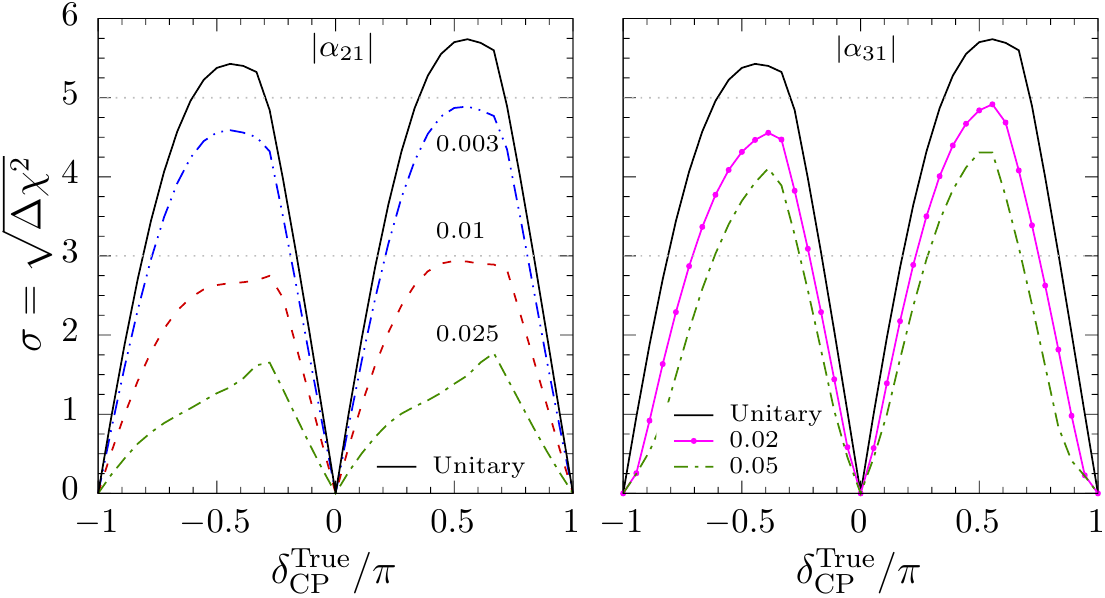}
\caption{DUNE sensitivity to CP violation for non-unitary neutrino
  mixing.  For comparison the black solid line shows the
  CP-sensitivity in DUNE, for the standard unitary case.  The reduced
  sensitivities for non-zero $\alpha_{21}$ ($\alpha_{31}$) are shown in the left (right) panel.  All undisplayed
  parameters have been marginalized over, including the non-unitarity
  phases as well as the diagonal parameters $\alpha_{ii}$.}
\label{fig:CPsens}
\end{figure}
In this section we analyze how DUNE sensitivity to the standard CP
violation is affected by the presence of non-unitarity.
To the oscillation parameters present in the standard unitary scenario
this analysis implies the addition of nine real parameters describing
the non-unitary mixing: the three real $\alpha_{ii}$ plus the three
complex non-diagonal $\alpha_{ij}$.
 
In order to simplify the analysis, we consider however only five
non-zero non-unitary parameters at a time: the three diagonal ones,
plus one of the non-diagonal ones, with its complex phase at a time. 
The resulting CP sensitivity in the presence of non-unitarity is shown
in Fig.~\ref{fig:CPsens}.  
As in the standard $\delta_{\text{CP}}$-sensitivity plot, the
CP-violation hypothesis is tested with respect to a CP-conserving
scenario \cite{Acciarri:2015uup}:
\begin{equation}\label{eq:chi2}
\Delta \chi^2 (\delta_\text{CP}^\text{true})=\text{Min}\left[\Delta \chi^2_{CP}(\delta_\text{CP}^\text{true},\delta_{\text{CP}}^\text{test}=0),\Delta \chi^2_{CP}(\delta_\text{CP}^\text{true},\delta_{\text{CP}}^\text{test}=\pi)\right],
\end{equation}
 The remaining standard $\vec{\lambda}$ as well as the non-unitarity parameters $\vec{\alpha} \equiv \{\alpha_{ii},\phi_{ij}\}$, which are included in both the simulated and reconstructed event rates in DUNE, $n(\vec{\lambda},\delta_{\text{CP}};\vec{\alpha})$ , have been marginalized over.
The left panel has been obtained for different values of
$|\alpha_{21}|$, while the right panel corresponds to the results of
the corresponding analysis performed for the  non-diagonal
non-unitarity parameter $\alpha_{31}$. 
One sees from the left panel that the sensitivity to the Dirac CP
phase decreases in the presence of non-unitarity with respect to the
standard `unitary' case, shown in the black-solid line.  
The remaining lines correspond to the non-unitary case with different
values for the $\alpha_{21}$ parameter, as indicated.  We have
selected three different benchmark values, the smaller one, $0.003$,
consistent with the upper part of Table~\ref{tab:bounds} and $0.010$ and
$0.025$ consistent with sensitivities displayed in the lower part of the
table, obtained from neutrino data only.
Even taking at face value the ``aggressive'' sensitivity
$|\alpha_{21}|=0.003$, the significance of a CP-violation measurement
decreases by $0.85\sigma$, compromising the possibility of testing any
range of values of the CP phase at $5\sigma$. 
For more conservative and reasonable choices $|\alpha_{21}|$ at the
1\% level one sees that the presence of non-unitarity precludes our
ability to probe CP violation at $3\sigma$ for nearly all of the
$\delta_{CP}$ range. 
One sees that probing maximal CP violating values $\pm \pi/2$ with
high significance in the presence of non-unitarity for `large'
$|\alpha_{21}|$ constitutes a big challenge for DUNE. 
In the right panel, we show the results of the same analysis for the
non-diagonal parameter $\alpha_{31}$.  As we discussed before, 
the impact of this parameter on the neutrino oscillation probabilities in DUNE 
is significantly less relevant in comparison with $\alpha_{21}$.
As a result, one can see from the figure that even if the fraction of CP
at $5\sigma$ is largely reduced respect to the unitary case, the reduction in the significance of CP tests 
is much smaller than for $|\alpha_{21}|$. This result holds for relatively large values of the parameter
compatible with the bounds in the lower part of  Table~\ref{tab:bounds},  namely $|\alpha_{31}|=0.05$.
The effect of the third non-diagonal parameter, $\alpha_{32}$, is not displayed in the figure. We have checked that it 
plays nearly no role in the analysis,  which confirms our discussion in
Section \ref{sec:non-unit-oscill}.
%

Here we note that Ref.~\cite{Dutta:2016vcc} has also discussed the
possibility of probing CP violation with T2K, NOvA and DUNE in the
presence of non-unitarity. Although we have a qualitative agreement in
the loss of CP-sensitivity due to the presence of non-unitarity, our
results show some quantitative differences. We ascribe these
discrepancies to the treatment of the DUNE simulation. Here we are
using the official description released by the DUNE Collaboration, and
we have validated our method against the official DUNE CP-sensitivity
result for the standard (unitary) oscillation analysis.

\subsection{DUNE sensitivity to non-unitary neutrino mixing}
\label{sec:DUNE-a21}

\begin{figure}[t!]
\includegraphics[width=0.8\textwidth,angle=0]{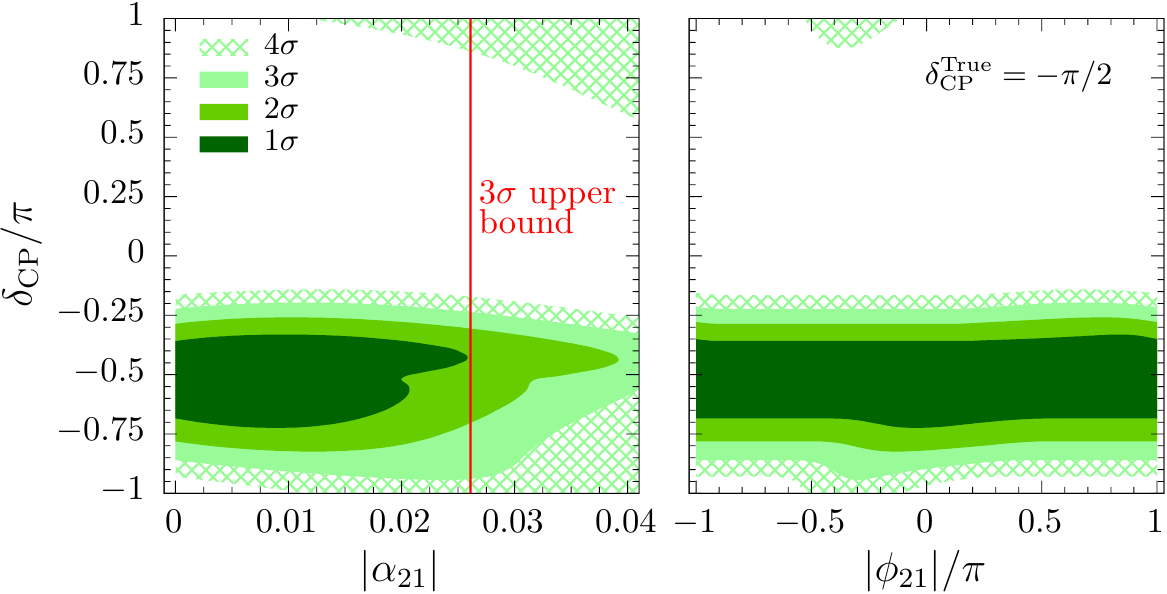}
\caption{Testing non-unitary neutrino mixing against the
  standard case, when only the
  off--diagonal parameter $\alpha_{21}$ is present.
  In the left (right) panel, the $\alpha_{21}$--$\delta_{CP}$
  ($\phi_{21}$--$\delta_{CP}$) allowed parameter space is shown. The 3$\sigma$ upper bound on $\alpha_{21}$
from neutrino data is indicated by a red line. This limit has also been included as a prior in the results 
  shown in the right panel. Here
  $\delta_{CP}^{\text{True}}$ is fixed to $-\pi/2$ and the additional
  undisplayed parameters have been marginalized over. The allowed
  regions, darkest to lightest, correspond to $1\sigma$ to $4 \sigma$ for 2 d.o.f., respectively.}
\label{fig:dcpsensitiv}
\end{figure}

In this section we analyze the potential of DUNE in constraining the
non-unitarity of the neutrino mixing matrix.
As we have discussed in Section~\ref{sec:prel-updat-bounds}, the most
robust and direct of these constraints come from neutrino oscillation
experiments and are not very strong.  Therefore, we wish to explore the
capability of DUNE in further constraining non-unitarity.
For this purpose we will focus in the analysis of the neutrino signal
at the DUNE far detector. The capability of the near detector will be
analyzed in the future. 
As we have discussed in the previous subsection, the parameter with
the most impact on the DUNE sensitivity to CP violation is $\alpha_{21}$.
As a result we will focus on $\alpha_{21}$ as the key parameter to be
constrained in order to characterize the loss of sensitivity in CP searches
at DUNE.
Following the usual procedure in analyzing the sensitivity of a
  given experiment to an unknown parameter (the non-unitary parameter
  $\alpha_{21}$ in this case), we have simulated DUNE events under the
  hypothesis of unitary mixing
  $n^{\text{true}}(\vec{\lambda},\delta_{\text{CP}}^\text{true})$.
  Afterwards, we have tried to reconstruct DUNE data in terms of the non-unitary neutrino mixing ansatz,
  $n^{\text{test}}(\vec{\lambda},\delta_{\text{CP}}^\text{test};\vec{\alpha})$.
 %
%
  It is worth noticing that the treatment of the non-unitarity here is
  different from the analysis performed in the previous subsection and
  therefore a direct comparison between the results presented in
  Figs. \ref{fig:CPsens} and \ref{fig:dcpsensitiv} is not
  straightforward.
For this analysis, the true value of the Dirac CP phase has been fixed to its current
preferred value, $\delta_{\text{CP}}^{\text{True}}=-\pi/2$. After
marginalizing over the diagonal non-unitary parameters and all the
oscillation parameters but $\delta_{\text{CP}}$, we obtained the
allowed parameter regions (at $1-4 \sigma$ for 2 d.o.f) shown in
Fig.~\ref{fig:dcpsensitiv}.
In the left panel, the allowed regions in the
$\delta_{\text{CP}}-|\alpha_{21}|$ plane show that DUNE is sensitive
to values of $|\alpha_{21}|$ at the percent level at $1\sigma$.
 As expected, the best fit point for the Dirac CP phase is equal to
 the assumed `true' value.
 However, for large enough values of $|\alpha_{21}|$, degenerate
 solutions around $\delta_{\text{CP}}= \pm \pi$ appear at higher
 C.L. 

\begin{figure}[t!]
\includegraphics[width=0.6\textwidth,angle=0]{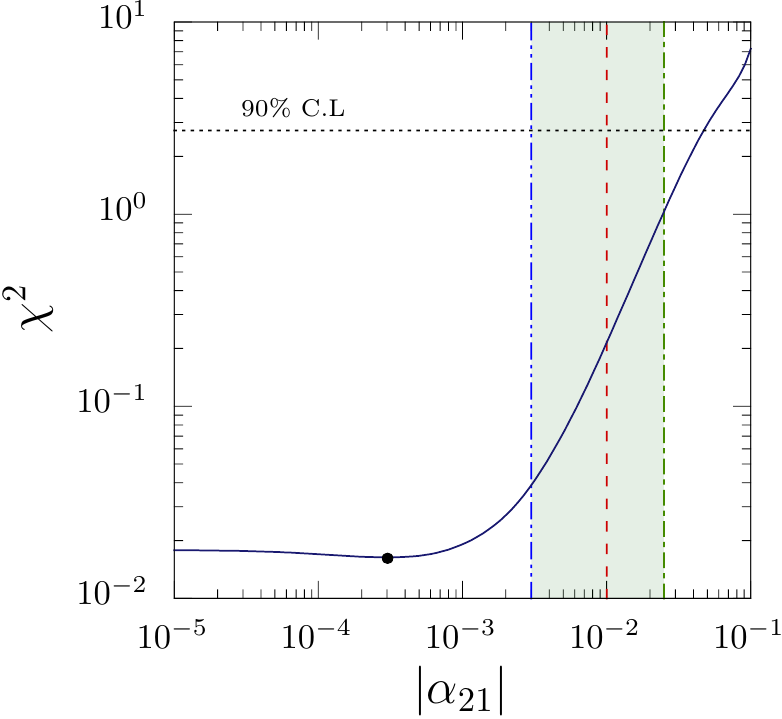}
\caption{DUNE sensitivity to $\alpha_{21}$ for arbitrary values of
  the Dirac CP phase. 
  The expected sensitivity 
  at $90\%$ of C.L.  is indicated by the horizontal
  line, while the three vertical lines correspond to the benchmark points considered in Fig.~\ref{fig:CPsens}. }
\label{fig:a21sensitiv}
\end{figure}

 Finally, we present an estimate of the absolute sensitivity of
  DUNE to the non-unitary parameter $\alpha_{21}$. In order to do
  this, we extended our previous analysis, considering all the
  possible values of $\delta_{\text{CP}}^{\text{True}}$ and
  marginalizing over $\delta_{\text{CP}}$ and $\phi_{21}$.
  Fig.~\ref{fig:a21sensitiv} shows the $\chi^2$ profile obtained as a
  function of $|\alpha_{21}|$ after marginalizing over all the
  remaining parameters, including $\delta_{\text{CP}}^{\text{True}}$.
  The best fit point, denoted by a black point in the figure, is
  obtained for $|\alpha_{21}| = 3\times 10^{-4}$. Nevertheless, the
  preference over the unitary hypothesis is not significant at all, as
  can be seen from the figure.
  The shaded band in Fig.~\ref{fig:a21sensitiv} indicates the three
  benchmark values of $\alpha_{21}$ used in the analysis of CP
  sensitivity in DUNE (see the left panel of Fig.~\ref{fig:CPsens}),
  while the horizontal dotted black line defines the parameter region
  allowed by DUNE at $90\%$ of C.L., corresponding to the limit
  $|\alpha_{21}|<0.046$.
 This bound is somewhat weaker than the constraints derived from
  neutrino oscillation searches, indicating that the analysis of
  long-baseline neutrino oscillations in DUNE is not expected to
  improve our current knowledge on the non-unitarity of the neutrino
  mixing matrix.
  However, it is worth mentioning that this constraint can also be
  regarded as independent and complementary to the bounds in
  Table~\ref{tab:bounds}.

\section{Conclusion and discussion}
\label{sec:conclusions}

    We have reviewed the existing limits on non-unitarity parameters,
    from weak universality considerations as well as from neutrino
    oscillation data. We have discussed the model-independent
    character of the latter constraints, since they only rely upon
    direct information from the neutrino sector, in contrast with the
    ones derived form charged lepton processes.
    We have developed in detail the formalism for neutrino propagation
    in matter in the presence of non-unitary neutrino mixing. In
    contrast to the standard unitary case, the neutral current
    potential contributes to the neutrino Hamiltonian in matter.
    Here we have focused our analysis on the case of the long-baseline
    neutrino experiment DUNE.
    First we have analyzed how the sensitivity to CP violation in DUNE
    can be affected by the presence of non-unitarity.
   We have found that DUNE's potential to probe CP violation is
      somewhat weakened, although not as much as one might have
      expected, see Fig.~\ref{fig:CPsens}. The reason for this, apart
      from the high statistics, is mainly the fact that the DUNE
      experiment is characterized by a relatively wide beam, compared
      with current experiments. This nice feature partly mitigates the
      ambiguities stressed in~\cite{Miranda:2016wdr}.
      Moreover, we have investigated how DUNE can probe neutrino
      properties beyond standard oscillations, such as the parameters
      characterizing non-unitarity, see Fig.~\ref{fig:a21sensitiv}.
      In this respect DUNE is not expected to perform better than
      previous short baseline oscillation searches at  
      NOMAD, CHORUS and NuTeV. This discouraging result is not surprising, as the
      sensitivity to non-unitarity comes mainly from probing the
      ``zero-distance effect'' and hence involves ``near'' detection.
      This could be improved within a setup of the type suggested in
      Ref.~\cite{Ge:2016xya}.

    Before closing, we mention two other recent related analyses. In
    Ref.~\cite{Blennow:2016jkn} the authors considered the effect of
    extra neutrino states in neutrino oscillations, focusing on the
    differences and similarities between the case in which these
    neutrinos are kinematically accessible (sterile neutrinos) or not
    (non-unitary mixing matrix)\footnote{The comparison between these two scenarios has also been explored in Ref.~\cite{Fong:2016yyh}.}. They choose the stronger
    model-dependent bounds, for which the effect of non-unitarity on
    the neutrino signal in DUNE is very small and hence the loss in CP
    sensitivity. 
      Finally, Ref.~\cite{Pas:2016qbg} suggested a novel possibility
      of discriminating sterile neutrinos and unitarity violation
      through CP violation.  

\acknowledgements

This work was funded by the Spanish grants FPA2014-58183-P, Multidark
CSD2009-00064, SEV-2014-0398 (MINECO) and PROMETEOII/2014/084
(Generalitat Valenciana).
MT is also supported by the grant GV2016-142 (Ayuda para Proyectos
Emergentes, Generalitat Valenciana) and by a Ram\'{o}n y Cajal
contract (MINECO).
OGM is supported by the CONACyT grant 166639 (Mexico).
DVF thanks the URA fellowship that allowed him to visit the theory
division at Fermilab where some part of this project was carried on.
DVF has been supported by the U.S. Department Of Energy under the
DE-SC0013632 and DE-SC0009973 contracts.

\bibliographystyle{apsrev} 
\providecommand{\url}[1]{\texttt{#1}}
\providecommand{\urlprefix}{URL }
\providecommand{\eprint}[2][]{\url{#2}}

\end{document}